\documentclass[manuscript]{aastex63}

\received{Received date }
\revised{Revised date}
\accepted{Accepted date }
\submitjournal{ApJ}

\shorttitle{Interferometric LOFAR observations of J1740+1000}
\shortauthors{Ro{\.z}ko et al.}
\graphicspath{{./}{figures/}}

\begin{document}

\title{The significance of low frequency interferometric observations for the GPS pulsar flux estimation: the case of J1740+1000}

\correspondingauthor{Karolina Ro{\.z}ko}
\email{k.rozko@ia.uz.zgora.pl}

\author[0000-0002-1756-9629]{K. Ro{\.z}ko}
\affiliation{Janusz Gil Institute of Astronomy\\
University of Zielona G\'ora \\
ul. Prof. Z. Szafrana 2, \\
65-516 Zielona G\'ora, Poland}

\author[0000-0001-9577-708X]{J. Kijak}
\affiliation{Janusz Gil Institute of Astronomy\\
University of Zielona G\'ora \\
ul. Prof. Z. Szafrana 2, \\
65-516 Zielona G\'ora, Poland}

\author[0000-0002-6280-2872]{K. Chy{\.z}y}
\affiliation{Astronomical Observatory\\
Jagiellonian University\\
ul. Orla 171\\
30-244 Krak\'ow, Poland}

\author[0000-0003-0513-9442]{W. Lewandowski}
\affiliation{Janusz Gil Institute of Astronomy\\
University of Zielona G\'ora \\
ul. Prof. Z. Szafrana 2, \\
65-516 Zielona G\'ora, Poland}

\author[0000-0001-5648-9069]{T. Shimwell}
\affiliation{ASTRON\\
Netherlands Institute for Radio Astronomy\\
Oude Hoogeveensedijk 4\\
7991 PD Dwingeloo, The Netherlands}

\author[0000-0002-7587-4779]{S. S. Sridhar}
\affiliation{ASTRON\\
Netherlands Institute for Radio Astronomy\\
Oude Hoogeveensedijk 4\\
7991 PD Dwingeloo, The Netherlands}

\author{M. Cury\l{}o}
\affiliation{Astronomical Observatory\\
The University of Warsaw\\
Al. Ujazdowskie 4\\
00-478 Warsaw, Poland}

\author[0000-0003-2812-6222]{A. Krankowski}
\affiliation{Space Radio Diagnostics Research Center\\
University of Warmia and Mazury in Olsztyn\\
ul. Michala Oczapowskiego 2\\
10-719 Olsztyn, Poland}

\author[0000-0003-0928-9616]{L. B\l{}aszkiewicz}
\affiliation{Space Radio Diagnostics Research Center\\
University of Warmia and Mazury in Olsztyn\\
ul. Michala Oczapowskiego 2\\
10-719 Olsztyn, Poland}

\nocollaboration{9}



\begin{abstract}
 In this paper we present recent Low Frequency Array (LOFAR) observations of the pulsar J1740+1000. We confirm that its
 spectrum has a turnover at 260~MHz, which is unusual for a typical pulsar. We argue that in this case interferometric imaging provides more accurate pulsar flux estimates than other, more traditional, means such as beamformed observations. We conclude that existing calibration and imaging techniques can be used for a more comprehensive study of the influence of the interstellar medium on the point-like sources at very low frequencies in the near future. 
\end{abstract}

\keywords{Radio pulsars (1353); Interstellar medium (847)}


\section{Introduction} \label{sec:intro}

The flux of the majority of pulsars increases with decreasing frequency at least down to about 100~MHz. Their spectra are well described by a single power-law function with the population average spectral index\footnote{Spectral index describes the dependence of flux density on frequency, if $S_{\nu}$  $\sim~\nu^{\alpha}$, then $\alpha$ is a spectral index.} close to -1.6 \citep{1995Lorimer,2017Jankowski}. However, for some pulsars low-frequency observations made by the standard technique (i.e. recording pulsars profiles, \citealt{2012Lorimer}) have been shown to be inaccurate at measuring the pulsar's flux. The main reason for that is the influence of the interstellar medium
(ISM), which introduces a scattering tail on the trailing side of the pulsar profile. In extreme cases the characteristic scattering time may approach or exceed pulse period, substantially hindering source detection. In even more extreme cases the pulsar's flux can be absorbed via free-free thermal absorption caused by an ionized medium in the vicinity of the pulsar (i.e. dense filaments in supernova remnants, bow-shock pulsar wind nebulae, or H II regions). The absorption becomes more severe at lower frequencies and this results in an observed turnover in the apparent pulsar spectrum and a peak flux at around 1 GHz \citep{2011KijakA,2011KijakB,2013Kijak,2017Kijak}. These sources were called the gigahertz-peaked spectra (GPS) pulsars and currently $\sim~30$ such objects are known \citep[see][and references therein]{2015Dembska,2016Basu,2018Basu,2017Kijak,2018Kijak,2017Jankowski}.

It is well known that some pulsars exhibit a genuine low-frequency turnover i.e. below 500 MHz \citep[see, for example,][and references therein]{1973Sieber,1996Malofeev,2003Kramer}. The most recent observational studies include those by \citet{2017Murphy}, \citet{2016Bilous}, \citet{2020Bilous}, and \citet{2020Bondonneau}. To date several mechanisms have been proposed to explain this phenomenon, as described in the review article by \citet{2002Sieber}.

The PSR~J1740$+$1000 spectrum was classified by \citealt{2014Dembska} as \textbf{a} GPS; however, the recent 150~MHz Low Frequency Array LOFAR beamformed measurements by \citet{2016Bilous} challenged the GPS interpretation. Their flux density estimate suggest that the PSR J1740+1000 spectrum can be described by a single power-law function. Other recent wide frequency (from 325 MHz to 5900 MHz) observations indicate a turnover at the frequency around 550~MHz, but the low-frequency $(<325~\mathrm{MHz})$ behavior still remains unclear \citep{2018Rozko}. 

Interferometric imaging is usually more accurate than beamforming techniques for the low-frequency flux estimation of GPS pulsars \citep{2016Basu,2017Kijak}. The most important advantage is that instrumental and atmospheric gain fluctuations can be corrected on very short time-scales by self-calibrating the interferometric data, which is not feasible for beamformed pulsar observations.

In this paper, we present new 120-168 MHz observations (project LC$9\_004$) of PSR J$1740+1000$ obtained with LOFAR (\citealt{2013vanHaarlem}). We utilised the multibeam capability of LOFAR and observed our target for 2 x 4 hr simultaneously with one of the LOFAR Two-Metre Sky Survey (LoTSS, \citealt{2017Shimwell}) fields. PSR J$1740+1000$ was also independently observed for 2 x 4 hr as part of LoTSS (pointing P$265+10$, proposal code LT$10\_010$). This LoTSS pointing is centred 0.8 degree from the target, which lies approximately at the 50\% level of the power primary beam. We use these observations to study the low-frequency behavior of PSR J$1740+1000$. 
With this example we also highlight the capability of interferometric imaging observations at low frequencies and advocate using this to study all pulsars where the flux density is strongly affected by the ISM. 

The outline of the paper is as follows. In the Section 2 we describe our observations and the calibration techniques we used. In Section 3 we compare the maps we obtained using different calibration procedures and also the different observations. In Section 4 we discuss the PSR J$1740+1000$ spectrum and the implications of our new measurements. In Section 5 we conclude.

\section{Observations and data reduction}\label{sec:obs_reduction}

\subsection{LOFAR Observations and Preprocessing}\label{subsec:obs_general}

We observed PSR~J1740+1000 using the LOFAR High Band Antenna (HBA) for four hours on both 2017 December 12 and 2018 February 1. Each observation was bracketed by short scans of flux density calibrators. In addition to these targeted observations, we made use of data from LoTSS. For an overview of LoTSS and a detailed description of the observation and processing strategy, we refer the reader to \cite{2017Shimwell}. Parameters related to the three observations are listed in Table~\ref{tab:general}.

The observational setup of all three runs were identical. The observing bandwidth covers the frequency range from 120 to 187~MHz and was split into 243 subbands (SBs), which were subdivided into 64 channels of 3~kHz width. The correlator integration time was set to 1~s. After the observation, radio frequency interference (RFI) excision was carried out on the high time and frequency resolution data before they were averaged in frequency to 16 channels per subband. RFI excision was carried out using \texttt{AOFlagger} \citep{offringa2010,offringa2012} and averaging was done with the New Default PreProcessing Pipeline \citep[\texttt{NDPPP}][]{vandiepen2018}. Only the averaged visibility data are stored in the LOFAR Long Term Archive (LTA)\footnote{\url{https://lta.lofar.eu}}.

LOFAR HBA data are first calibrated to correct for direction independent effects such as clock offsets and bandpass, which can be derived from calibrator observations and transferred to the target field \citep[see, e.g.][]{2019DeGasperin}. After the direction independent calibration is performed the image fidelity is still low and direction-dependent calibration is needed to correct for ionospheric distortions that are severe at low frequencies. Currently, there are two widely used routines to perform direction-dependent calibration on LOFAR imaging data, namely \texttt{Factor} \citep{2016vanWeeren}\footnote{\url{https://github.com/lofar-astron/factor}} and \texttt{DDF}-pipeline \citep{2019Shimwell}\footnote{\url{https://github.com/mhardcastle/ddf-pipeline}}. In this project, we process our data using both techniques as summarized below.

\subsection{\texttt{Factor} data analysis}\label{subsec:calib1}

We calibrated and imaged data from the targeted observation using \texttt{Factor} \citep{2016vanWeeren,2016Williams}. In our \texttt{Factor} processing we divided the field of view into multiple facets such that each facet has at least a point source with integrated flux density greater than 0.4~Jy. For each facet, we derived calibration solutions by self-calibrating a small region around the facet calibrator and applied the solutions to the other sources in that facet. Once calibrated, we imaged the calibrated visibilities from the facet containing PSR~J1740+1000 using the wideband deconvolution algorithm available in \texttt{WSClean} \citep{offringa2014}.

\subsection{\texttt{KillMS} and \texttt{DDFacet} data analysis}\label{subsec:calib2}

Both the targeted data set and the LoTSS data set were processed using the latest version of the LoTSS pipeline (as used for LoTSS-DR2). This pipeline utilizes \texttt{DDFacet} \citep{2018Tasse} and \texttt{KillMS} \citep{2014Tasse,2015Smirnov,2020Tasse}\footnote{This pipeline is also summarized in Sec.~2.3.3 of \citet{2019Shimwell}}. To further enhance the image fidelity in the region of the pulsar, we used the direction-dependent calibration solutions to remove all sources away from the target and then performed an additional self-calibration cycle on this region. This postprocessing, which is described in detail in \citet{2020vanWeeren}, allows us to optimize the calibration toward the pulsar and to produce a data set calibrated in the direction of the target that can be easily reimaged. 

\begin{table}
	\centering
	\caption{Observations and Imaging Parameters}
	\label{tab:general}
	\begin{tabular}{lccc} 
		\hline
		Observation Type &  Targeted  & Targeted  & LoTSS Pointing\\
		Calibration Method & \texttt{Factor} Pipeline &  \texttt{DDF}-pipeline & \texttt{DDF}-pipeline\\
		\hline
		Observation Date & 2017 Dec 12 (4 hr) & 2017 Dec 12 (4 hr) &  2018 Sep 24 \\
		                 &  2018 Feb 01 (4 hr) & 2018 Feb 01 (4 hr)& \\
		Map resolution [arcsec] & $16.5$ x $6.2$ & $6.0$ x $6.0$ & $12.33$ x $5.00$\\
		         rms $[\mathrm{mJy}/\mathrm{beam}]$ & 0.27 & 0.25 & 0.47 \\
        Integral intensity $[\mathrm{mJy}]$ &  $2.85 \pm 0.98$  &  $3.42 \pm 1.65$  &  $3.45 \pm 0.98$ \\

		\hline
	\end{tabular}
\end{table}

\section{Image analysis}
In this section we compare three maps: two from our targeted observations (their resolution is $16.5$ x $6.2$ arcsec and $6$ x $6 $ arcsec, respectively), and one from LoTSS observations (its resolution is $12.33$ x $5.0$ arcsec). 

The flux density values and the rms of the maps were measured using the The Astronomical Image Processing System (\texttt{AIPS}) software tasks JMFIT and IMSTAT, respectively. Even though the pulsar is point-like source we find a discrepancy between the peak and integral intensity measurements that is likely caused by residual ionospheric phase errors \citep{2017Shimwell} - we only use the integrated flux values for further analysis. 

To ensure an accurate flux scale of our LOFAR images the integrated flux densities of 60 bright, compact sources in our field of view were cross-checked with TGSS-ADR1 \citep{2017Intema}. We found that the median ratio of the integrated \texttt{DDF}-pipeline flux densities to the integrated TGSS-ADR1 flux densities is 1.09 (1.22 for LoTSS pointing observations), which lies within the typical range for LOFAR observations \citep[see, e.g][]{2019Shimwell}. \citealt{2019Shimwell} suggested that a conservative uncertainty of 20\% should be used on the LoTSS-DR1 integrated flux density measurements, but the dominant error in a flux density estimate comes from the signal-to-noise factor. As is seen in Table \ref{tab:general} our flux density errors are much bigger than proposed survey uncertainty, which is unsurprising since our source is very weak at 150~MHz and it was close to the detection threshold.

More surprisingly, the median ratio of the integrated LoTTS-\texttt{Factor} flux densities to the integrated TGSS-ADR1 flux densities is 0.72. This discrepancy is consistent with the PSR J1740+1000 \texttt{Factor} pipeline flux density estimate being lower than the flux density estimates obtained with \texttt{DDF}-pipeline. 

\begin{figure}[ht!]
\plottwo{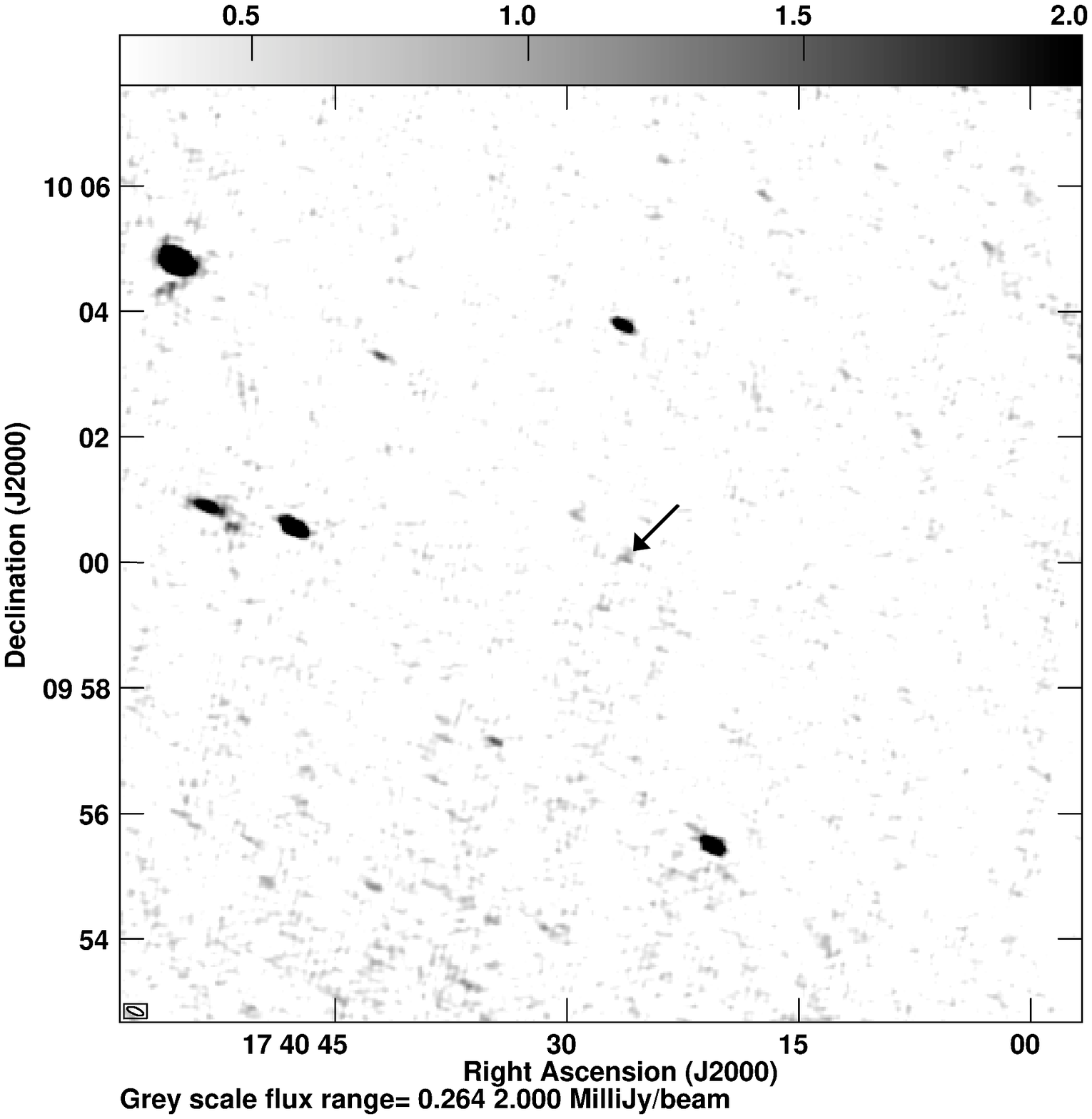}{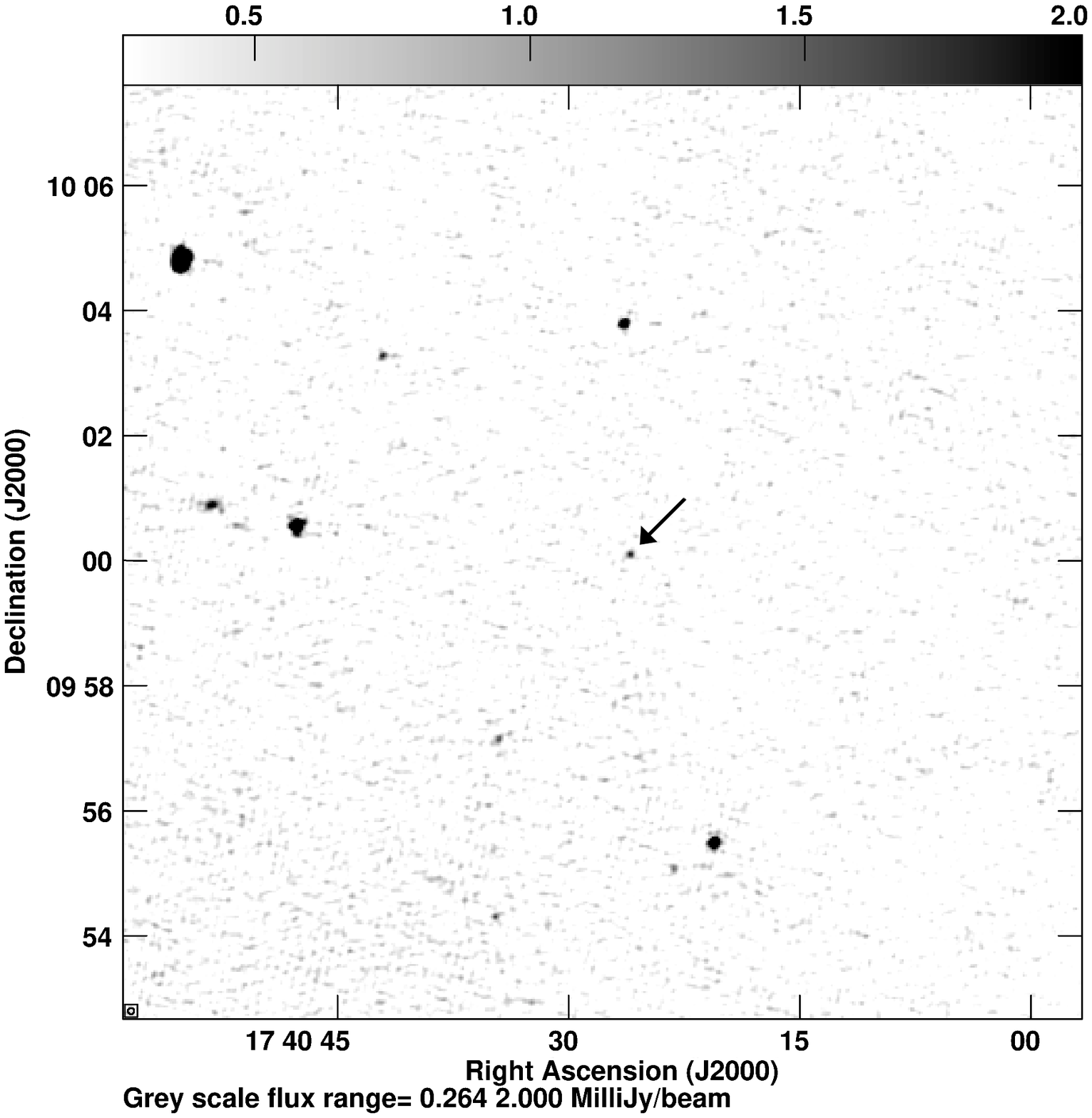}
\caption{Grayscale images showing the total intensity maps centered at the region around PSR's J$1740+1000$ position (the black arrow marks the pulsar position). Both maps were obtained from the targeted observations. Left: results from the \texttt{Factor} pipeline. Right: results from \texttt{DDF}-pipeline. The resolution of the images and the noise level around the pulsar position are listed in the Table \ref{tab:general}.   \label{fig:targeted_maps}}
\end{figure}

The map obtained from the \texttt{Factor} pipeline is shown in the left panel of Fig. \ref{fig:targeted_maps}. The noise level is similar to that obtained with \texttt{DDF}-pipeline. However, the detection of the pulsar in the \texttt{Factor} map was quite challenging but the detection is clear in the higher quality \texttt{DDF}-pipeline map (Fig. \ref{fig:targeted_maps} right panel) where the measured flux is $3.42 \pm 1.65$ mJy (see also Table \ref{tab:general}).   
  
\begin{figure}[ht!]
\centering
\includegraphics[width=0.45\textwidth]{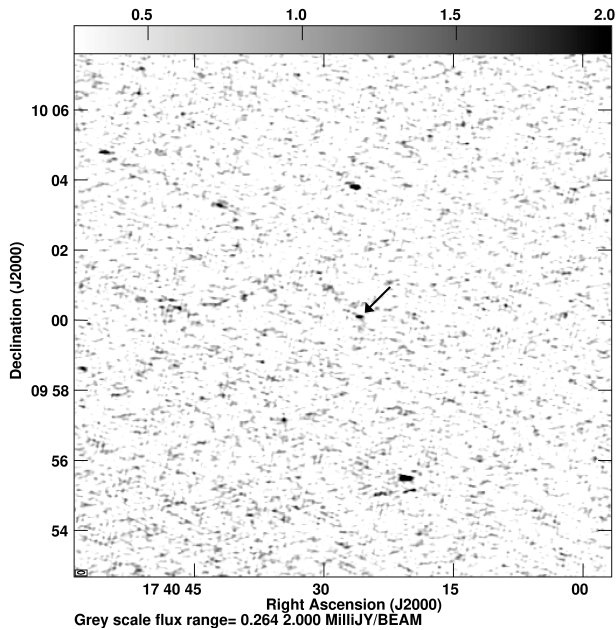}
\caption{Grayscale image showing the total intensity map of the LoTSS data centered on PSR J$1740+1000$ (the black arrow marks the pulsar position). The map resolution and noise around the pulsar position are listed in the Table \ref{tab:general}.
\label{fig:ponting_maps}}
\end{figure}
 
In the LoTSS observation (Fig.~\ref{fig:ponting_maps}) the target was $\sim 0.8$ degrees away from the pointing center, which corresponds to 50\% of the primary beamwidth from the center, but this should not have any effect on our results besides the increased noise. Nevertheless, from this map we made an independent measurement of pulsar flux that agreed with our targeted observation.  

\section{PSR J$1740+1000$ spectrum analysis}\label{Sec4}

\begin{figure}[ht!]
\plotone{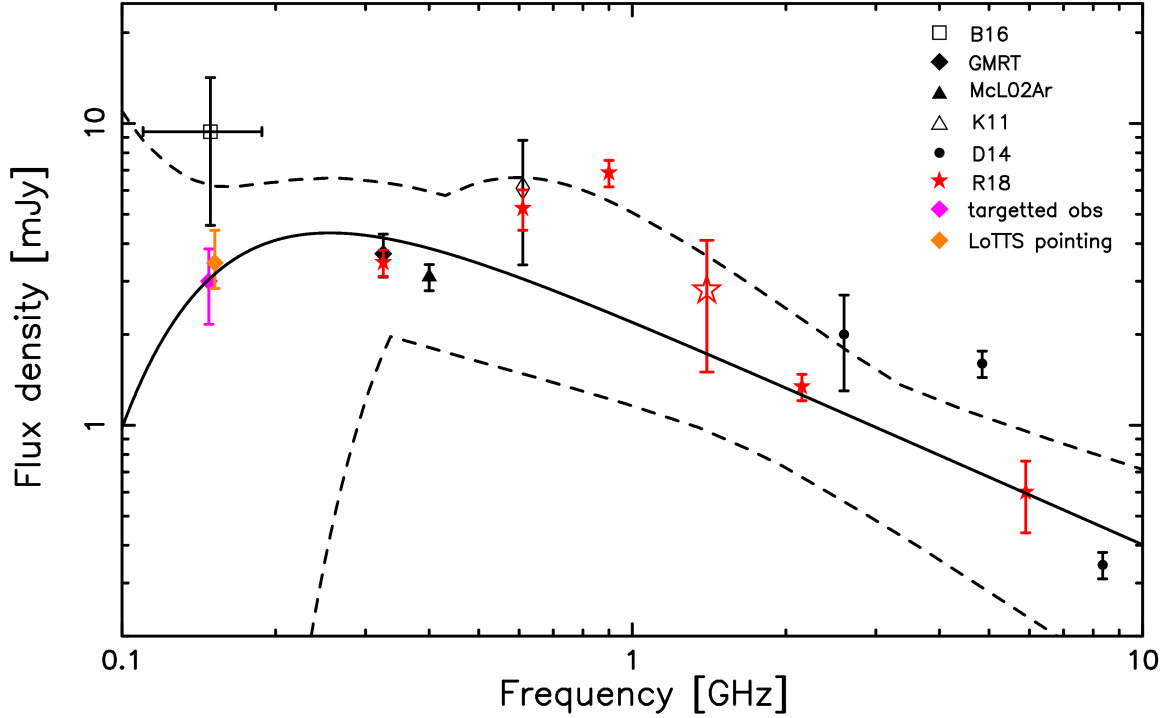}
\caption{The pulsar spectrum with a free-free thermal absorption model fitted to all available flux density measurements. The open red star denotes the flux density value that was averaged over 1--2~GHz observations. The two values at 150~MHz were slightly displaced in frequency in order to be more visible in the plot. The dashed lines correspond to $1 \sigma$ errors of the model. The values of the fitted parameters are presented in the Sec.~\ref{Sec4}. The acronyms point to the following publications: B16 -- \citet{2016Bilous}, GMRT -- \citet{2018Rozko}, McL02Ar -- \citet{2002McLaughlin}, K11 -- \citep{2011KijakB}, D14 -- \citet{2014Dembska}, R18 -- \citet{2018Rozko}. 
\label{fig:spectrum}}
\end{figure}

The pulsar J$1740+1000$ was discovered by \citealt{2000McLaughlin} using the Arecibo telescope. It is a young pulsar with a period of 154~ms that is located at a relatively large distance from the galactic plane, along a line of sight that includes the North Polar Spur (Loop I) and the Gould Belt, which is an expanding disk of gas and young stars \citep{2002McLaughlin}. This pulsar also hosts an unconfirmed Pulsar Wind Nebulae \citep{2008Kargaltsev,2010Kargaltsev}. Due to its location on the sky, diffractive interstellar scintillation makes flux measurements challenging and significantly different measurements have been made, especially at frequencies between 1 and 2~GHz. Figure \ref{fig:spectrum} shows all flux measurements we have found in the literature together with two new values: the weighted arithmetic mean flux value obtained from our two different processing methods used on the targeted observations, and the independent flux measurement from the LoTSS pointing.

 Generally, two observational techniques are used to estimate pulsar flux density: beamforming or single dish observations and interferometric imaging observations. For the beamforming or single dish observations the pulsar flux density is generally estimated from the pulsar mean profile that is accumulated over many pulses \citep{2012Lorimer}. The flux is determined by comparing the on pulse and off pulse emission levels. There are different ways of determining the baseline level that allow the measured on and off emission ratio to be translated into physical units. One method is to record the flux calibrator before and/or after pulsar observations. Another method uses the radiometer equation. For the interferometric observations the standard approach is to observe a flux calibrator before and at the end of each observational session for the data to be calibrated so that the intensity can be measured from the map. In case of the LOFAR this flux calibration is more complicated due to inaccuracies in the beam model \citep[e.g.][]{2019Shimwell}. Observational details and the flux density estimation methods for previously published  PSR~J1740$+$1000 studies can be found in \citealt{2018Rozko}.

PSR~J1740$+$1000 flux density values were estimated by different observational methods, which have different advantages and disadvantages. \citealt{2015Dembska} showed, based on a study of four pulsars that were heavily affected by scattering, that the flux measurements obtained using beamformed methods be underestimated in comparison with flux density measurements from interferometric imaging observations. However, in cases where the scattering time is very small or negligible both flux density estimates are consistent within measurement errors \citep{2015Dembska,2016Basu}. The mean profiles of PSR~J1740$+$1000 published in the appendix of \citealt{2018Rozko} showed that with decreasing frequency the mean profile weakens but does not significantly broaden, i.e. no scattering tail appears on the  trailing side of pulsar profile. This implies that both methods of the flux density estimation at frequencies exceeding $\sim 300$~MHz should give the same result. 

Regardless of the method, the pulsar flux density measurements should be affected by the scintillation-driven flux variation in exactly the same way. The typical time scale for the diffractive scintillaton at LOFAR frequencies is around a few minutes. For our 8 hr LOFAR observations we thus average over these possible flux variations. In the case of refractive scintillations the time variation scale can vary from days to months. Targeted and LoTSS observations were separated around a half a year, so any significant refractive scintillations should manifests in different flux density levels in both observations, but instead we observed very similar flux density values.

To check the GPS interpretation of the pulsar spectrum we fitted a free-free thermal absorption model to all available flux density values. We follow the approach of \citet{2015Lewandowski}, which is similar to that employed by \citet{2016Rajwade}), and use the following formula of the flux ($S_{\nu}$) at any frequency ($\nu$):
\begin{equation}
 S_{\nu} = \mathrm{A} \left( \frac{\nu}{10} \right)^{\alpha} e^{-B\nu^{-2.1}},
\end{equation}
where $A$ is the pulsar intrinsic flux at 10 GHz, $\alpha$ is the pulsar intrinsic spectral index, and $\nu$ is a frequency in GHz. The parameter $B$ is defined as:
\begin{equation}
B = 0.08235 \times \left(\frac{T_{\mathrm{e}}}{\mathrm{K}} \right)^{-1.35}~\left(\frac{\mathrm{EM}}{\mathrm{pc}~\mathrm{cm}^{-6}}\right),
\end{equation}
where $T_{\mathrm{e}}$ is the electron temperature and EM is the emission measure. In the fitting procedure $A$, $\alpha$ and $B$ were free parameters, and ${\nu}_{\mathrm{p}}$ is the frequency corresponding to the maximum pulsar flux. To fit the data we used the Levenberg-Marquardt nonlinear least squares algorithm (\citealt{1944Levenberg}, \citealt{1963Marquardt}) and estimated the errors using $\chi^2$ mapping \citep{1996Press}. 

The fitted parameters are $A = 0.40^{+0.31}_{-0.27}$, $B = 0.020^{+0.23}_{-0.020}$, $\alpha  = -0.75^{+0.36}_{-0.72}$, $\chi^2 = 2.01$, and $\nu_{\mathrm{p}} = 0.26$. Figure~\ref{fig:spectrum} shows the pulsar spectra with the fitted model and $1 \sigma$ errors in this. According to new estimated parameters the peak frequency shifted toward lower frequency ($\nu_{\mathrm{p}} = 260$ MHz), which is lower than that deduced in \citealt{2018Rozko} ($\nu_{\mathrm{p}} = 540$ MHz). Our result also confirms that the spectrum of PSR~J1740$+$1000 is not a power law. 

We used the pulsar dispersion measure (DM) to constrain the electron density and temperature of the absorber \citep[see][and references therein]{2018Rozko}. Assuming that half of the DM is contributed by the absorber  we calculated the EM for three different absorber cases: a dense supernova remnant filament (with size equal $0.1$~pc), a pulsar wind nebula (with size equal to $1.0$~pc), and a warm H II region (with size equal $10.0$~pc). For each case from the fitted value of parameter $B$ we obtained constraints on the electron temperature: $460^{+5280}_{-460}$ K (for 0.1 pc), $113^{+960}_{-113}$ K (for 1 pc), and $20^{+174}_{-20}$ K (for 10 pc). The derived temperatures are a little larger than those previously published, but they are consistent within the error, thus we still believe that the most probable absorber is a partially ionized small molecular cloud along the line of sight of the pulsar, because the derived electron temperature for the 10 pc absorber is too low for a H II region, and a presence of the PWN around the pulsars is still unconfirmed.  

Our flux density estimate at 150 MHz has fewer uncertainties than the flux density value reported by \citet{2016Bilous}. The main reason is that the PSR J1740+1000 profile is very weak and significantly broadened at 150 MHz (see Fig. C.8. in \citealt{2016Bilous}) which makes the task of finding the correct baseline level very difficult - their reported signal-to-noise ratio is to 6 but the flux density error is larger than 50\%. 

For the reasons stated above we believe that the interferometric method is better suited for low-frequency flux density measurements of this pulsar, as it is independent of the pulse broadening. However, our flux density estimate suggests that the simplest free-free absorption model with a single absorbing medium may be insufficient to explain observed spectrum shape. 

One possibility may be that with interferometric imaging observations we are measuring not only the pulsar flux, but also some additional flux density that comes from the pulsar wind nebula around the pulsar. If that was the case one can expect that while the pulsar flux would continue to sharply decrease at lower frequencies (due to effects such as thermal absorption), the total flux may behave differently because of the emission from the nebula itself. However, in this scenario the emission from the pulsar wind nebulae may experience synchrotron self-absorption at low frequencies. To check this hypothesis we need more measurements in the low-frequency part of the spectrum and a more complex model where synchrotron self-absorption is fitted together with thermal absorption.  

Another possibility is that there are two absorbers along the line of sight that can also explain the observed spectrum shape. This is not unlikely as J1740+1000 lies behind the North Polar Spur Region, and we need additional knowledge of structures along the line of sight to the  pulsar to ascertain if this is the case. Further exploration of the multicomponent free-free absorption model is beyond the scope of this paper and requires additional measurements below 300 MHz. It is worth mentioning that more complex spectra with free-free thermal absorption have been observed like the radio galaxies with two turnovers \citep[see, e.g.][and references therein]{2017Callingham}.

The free-free thermal absorption in the pulsar magnetosphere was proposed to explain turnovers at low frequencies (i.e. around 100~MHz) by \citet{1979Malov}. Today we know that the pulsar radio emission is coherent and their magnetosphere is filled with highly relativistic electrons which should not absorb their emission \citep{2017Mitra}. We cannot be sure that the observed total flux comes only from the coherent mechanism: it could be the sum of the pulsed and unpulsed emission \citep{2011Basu}; however, we believe that this is very unlikely to be the case.

The case of PSR~J1740$+$1000 shows that the current techniques to calibrate imaging data open up the possibility of conducting a comprehensive study of the spectra for pulsars that are GPS candidates or are heavily affected by scattering at low frequencies. The interferometric observations are crucial for these studies as they are not impacted by scattering.
In some cases beamformed observations may suffer from scattering that mimicks a turnover feature (see, i.e. the case of PSR B1815-14 presented in \citealt{2015Dembska}). In such cases the interferometry method could be the only way to estimate pulsar flux density. 

\section{Summary}

In this paper we presented the results of the LOFAR interferometric imaging observations of the PSR~J$1740+1000$. We have shown that two independent LOFAR observations (i.e. targeted and from LoTSS pointing) confirmed a low flux density at 150 MHz. Due to its relatively low peak frequency ($\nu_{\mathrm{p}} = 260$ MHz) obtained from the model it should not be classified as a typical GPS pulsar. However,  flux measurements themselves still show a maximum flux around 600~MHz. We believe that in this case the mechanism responsible for the low low-frequency flux density may be still similar to the mechanism responsible for turnovers of GPS pulsars.  

We claim that measuring fluxes from interferometric imaging data is important for the GPS pulsar population in general and also for other pulsars that exhibit low-frequency turnovers (below 500 MHz). While this is not the case of PSR~J$1740+1000$, for other GPS candidates it could help to distinguish the true turnover from the "artificial" flux density drop that may result from the strong pulse broadening caused by the scattering. Moreover, the existing calibration techniques should make interferometric pulsar observations possible at frequencies below 150 MHz, which will in turn allow for a more comprehensive study of the influence of the ISM at very low frequencies \citep{2018DeGasparin}.  

Our study shows that in the case of PSR~J$1740+1000$ the simple free-free absorption model seems to be insufficient to describe the whole spectrum shape and a more complex model may be required. 

We analyzed the LOFAR data with two different direction-dependent imaging pipelines, \texttt{DDF}-pipeline and \texttt{Factor}. In  this particular case our target was detected at higher significance in the \texttt{DDF}-pipeline processed data, however, a detailed comparison between the pipelines is beyond the scope of this paper.

\acknowledgments
We thank the anonymous referee for the comments that helped improve
the paper. This paper is based on data obtained from facilities of the International LOFAR Telescope (ILT) under project code LC$9\_004$. LOFAR (van Haarlem et al. 2013) is the Low Frequency Array designed and constructed by ASTRON. It has observing, data processing, and data storage facilities in several countries that are owned by various parties (each with their own funding sources) and that are collectively operated by the ILT foundation under a joint scientific policy. The ILT resources have benefited from the following recent major funding sources: CNRS-INSU, Observatoire de Paris and Université d'Orléans, France; BMBF, MIWF-NRW, MPG, Germany; Science Foundation Ireland (SFI), Department of Business, Enterprise and Innovation (DBEI), Ireland; NWO, The Netherlands; The Science and Technology Facilities Council, UK; Ministry of Science and Higher Education, Poland. The authors of the Polish scientific institutions thank the Ministry of Science and Higher Education (MSHE), Poland for granting funds for the Polish contribution to the International LOFAR Telescope (MSHE decision No. DIR/WK/2016/2017/05-1)" and for maintenance of the LOFAR PL-612 Baldy, LOFAR PL-611 stations (MSHE decisions: No. 59/E-383/SPUB/SP/2019.1 and No. 46/E-383/SPUB/SP/2019, respectively). This research was partially supported by the grant 2018/29/B/ST9/02569 of the Polish National Science Centre.

%


\bibliography{2020_Rozko_et_al_rev}{}
\bibliographystyle{aasjournal}



\end{document}